%% file: main.tex
\documentclass[conference]{IEEEtran}
\usepackage[utf8]{inputenc}
\IEEEoverridecommandlockouts

\usepackage{subfiles}
\usepackage{svg}

\usepackage[hidelinks]{hyperref} 

\usepackage{footnote}

\usepackage{amssymb}

\usepackage{multicol}
\usepackage{multirow}
\usepackage{makecell}

\usepackage{tikz}
\usetikzlibrary[fit,positioning,topaths,calc]
\usepackage[caption=false]{subfig}

\usepackage{listings}

\title{A Framework to Assess Knowledge Graphs Accountability
\thanks{This work is supported by the ANR \href{https://dekalog.univ-nantes.fr}{DeKaloG} (Decentralized Knowledge Graphs) project, ANR-19-CE23-0014, CE23 - Intelligence artificielle.}}

\author{\IEEEauthorblockN{Jennie Andersen}
\IEEEauthorblockA{\textit{Univ. Lyon, INSA Lyon, CNRS,} \\
\textit{UCBL, LIRIS, UMR5205} \\
Villeurbanne, France \\
jennie.andersen@insa-lyon.fr}
\and
\IEEEauthorblockN{Sylvie Cazalens}
\IEEEauthorblockA{\textit{Univ. Lyon, INSA Lyon, CNRS,} \\
\textit{UCBL, LIRIS, UMR5205} \\
Villeurbanne, France \\
sylvie.cazalens@insa-lyon.fr}
\and
\IEEEauthorblockN{Philippe Lamarre}
\IEEEauthorblockA{\textit{Univ. Lyon, INSA Lyon, CNRS,} \\
\textit{UCBL, LIRIS, UMR5205} \\
Villeurbanne, France \\
philippe.lamarre@insa-lyon.fr}
\and
\IEEEauthorblockN{Pierre Maillot}
\IEEEauthorblockA{\textit{Univ. Cote d’Azur,} \\
\textit{Inria, CNRS, I3S} \\
Sophia Antpolis, France \\
pierre.maillot@inria.fr}
}

\begin{document}

\maketitle

\begin{abstract}
Knowledge Graphs (KGs), and Linked Open Data in particular, enable the generation and exchange of more and more information on the Web.
In order to use and reuse these data properly, the presence of accountability information is essential.
Accountability requires specific and accurate information about people's responsibilities and actions.
In this article, we define KGAcc, a framework dedicated to the assessment of RDF graphs accountability. It consists of accountability requirements and a measure of accountability for KGs. Then, we evaluate KGs from the LOD cloud and describe the results obtained.
Finally, we compare our approach with data quality and FAIR assessment frameworks to highlight the differences.
\end{abstract}


\section{Introduction}
\label{sec:intro}
Knowledge Graphs (KGs), and Linked Open Data in particular, enable the generation and exchange of more and more information on the web. This abundance of easily accessible data on the web offers many opportunities for researchers, companies or ordinary citizens. However, in order to share and use these data properly and legally, it is important to know some information about a knowledge graph, such as for what purpose it was created, by whom, etc.

Among the meta-information that contributes to the correct use of KGs, we focus on dataset accountability.
It requires to provide information about actions on the dataset, ``descriptive information and information on the people responsible for it''~\cite{oppold2020accountable}.
As a concept very close to transparency~\cite{weitzner2008information}, it requires information to be easily accessible~\cite{wyatt2018many}. For the semantic web in particular, where software agents are particularly present, meta-information about KGs ``needs to be available in a machine-readable format''~\cite{farber2018linked}. These two things combined, we consider that this information should be present and searchable within the data of the KG itself.

Consider the GDPR (General Data Protection Regulation) for instance, that aims to protect personal data. To do so, Article 17 about the right to be forgotten states that the ``subject shall have the right to obtain from the controller the erasure of personal data concerning him or her''\footnote{\url{https://eur-lex.europa.eu/eli/reg/2016/679/2016-05-04}}
as soon as it does not fall under the right of freedom of expression and information. Therefore, every KG holding personal information, such as Wikidata, should provide contact information of this controller, i.e. a person responsible for the data, and ideally allow users to access it directly via its SPARQL endpoint.
As another example, to avoid misinterpretation and to improve the (re)use of the data, it is often necessary to know for what purpose they were created, and for whom the data are intended. 
For instance, in some database mainly dedicated to teaching purposes, such as the MONDIAL Database\footnote{\url{https://www.semwebtech.org/mondial/10/}}, some inaccuracies can be tolerated (or even desired). However, it cannot be reused without precaution for other purposes.
Therefore, it should indicate its intended audience or its expected usage. However, when querying its SPARQL endpoint, this information is not available.
Accountability ensures that this kind of information of major interest is effectively available.
Several studies are already looking for meta-information, either as some particular aspects of data quality~\cite{farber2018linked,debattista2018evaluating,zaveri2016quality}, or as some requirements of the FAIR principles (Findability, Accessibility, Interoperability, Reproducibility)~\cite{amdouni2022faire,devaraju2021automated,rosnetfair,wilkinson2016fair}.
Yet, they neither take into account the information used in the two previous examples nor several other accountability information.
This highlights the importance of accountability as a specific and distinct characteristic of RDF datasets and the importance of their evaluation w.r.t. this aspect.

Hence, in this paper, we propose a new framework, KGAcc, dedicated to the assessment of RDF graphs accountability. It consists of organized accountability requirements and a measure of accountability.
We experiment it on many KGs offering a publicly available SPARQL endpoint.
Our accountability measure gives an indication of the accountability of KGs to dataset users and providers. It aims to guide users in their choice of one KG rather than another and to help providers to identify ways to improve their datasets.

To define such a measure, several questions arise, such as what meta-information is required? How to evaluate heterogeneous KGs?
First, to define requirements, we rely on  the LiQuID metadata model which focuses on dataset accountability~\cite{oppold2020accountable} in general. It provides an explicit list of accountability requirements expressed in natural language. The problem, then, is to adapt this model to the specificities of knowledge graphs and to define the requirements in terms of SPARQL queries.
To evaluate the KGs, we use the SPARQL-based test suite of the IndeGx framework~\cite{maillot2023indegx}.
We observe that most of them do not provide any easily accessible accountability information. However, as some KGs do answer some questions, it shows that our demand is reasonable and that KGs have a lot of room for improvement.
In addition, to illustrate the specificities of this measure, we compare it with several assessment frameworks for data quality and FAIRness.

The rest of the article is organized as follows.
Section~\ref{sec:related_work} describes the state of the art.
We define the KGAcc framework in Section~\ref{sec:definition}. Then, Section \ref{sec:results} is devoted to the description of the methodology for evaluating RDF graphs and the results obtained. We then compare our accountability measure with the existing assessments of knowledge graphs in Section \ref{sec:comparison}. Finally, we conclude in the last section.

\section{Related Work}
\label{sec:related_work}
Generally speaking, accountability requires that there is sufficient information to describe the data~\cite{oppold2020accountable}, the actions on the data, from its creation~\cite{rowe2009assessing} to its use~\cite{weitzner2008information}, and the people responsible for these data as well as these actions~\cite{oppold2020accountable,weitzner2008information}.
It may concern different levels of the information system, such as information accountability~\cite{weitzner2008information,rowe2009assessing}, systems~\cite{naja2021semantic}, and dataset accountability~\cite{oppold2020accountable}. To evaluate the accountability of knowledge graphs, we focus on this latter point.
The accountability of knowledge graphs may be considered as part of their data quality, in the broad sense. These last years, several studies have highlighted the many facets of the notion~\cite{farber2018linked,debattista2018evaluating,zaveri2016quality}. In addition, general monitoring tools such as SPARQLES~\cite{vandenbussche2017sparqles} and YummyData~\cite{yamamoto2018yummydata} have been proposed, enabling to assess and draw profiles of SPARQL endpoints.

As a matter of fact, measuring the accountability of KGs is a special case of assessing metadata completeness, which is defined as ``the degree to which metadata properties and values are not missing in a dataset for a given task''~\cite{issa2021knowledge}.
Many works consider the presence of meta-information to evaluate knowledge graphs.
Studies about the data quality of KGs~\cite{farber2018linked,debattista2018evaluating,zaveri2016quality} include many different metrics, among which a few focus on meta-information. For instance, provenance information is required by a metric on trustworthiness.
The FAIR principles~\cite{wilkinson2016fair} are also interested in meta-information.
One of the principles of findability states that ``data [must be] described with rich metadata'', and reusability requires that ``meta(data) are richly described with a plurality of accurate and relevant attributes'', including a license, and provenance information.
Therefore, the required meta-information may overlap between accountability, data quality and FAIRness while having their own specificities.
Because of the high variability of the actual implementations of these metrics and principles, we confront them with our own requirements at the scale of the RDF properties in section~\ref{sec:comparison}.

In order to define the KGAcc framework to measure the accountability of KG, we base our work on the LiQuID metadata model~\cite{oppold2020accountable} which considers datasets in general. It offers a way for datasets to represent accountability meta-information throughout their life cycle.
The model has been validated based on a real-world workload that relies on existing regulations (such as the GDPR) and an expert survey. To our knowledge, it is the only one to provide such a precise and explicit list of accountability requirements, presented in the form of questions that describe the model.
Our framework adapts the hierarchy and the associated questions of LiQuID, taking into account the expressiveness of the most common vocabularies. It provides the requirements as a set of SPARQL queries corresponding to the questions. Our very first experiments are shortly reported in~\cite{andersen2023accountability}. The work described in this paper relies on (\textit{i}) new experiments that enable to distinguish between each dataset of a SPARQL endpoint, and \textit{ii}) improved queries, taking into account more vocabularies. In addition, we provide a thorough comparison with other evaluation  frameworks~\cite{farber2018linked,debattista2018evaluating,zaveri2016quality,amdouni2022faire,rosnetfair}.

Finally, in order to conduct our experiments and to query KGs with our own set of queries, several frameworks can be used.
Luzzu~\cite{debattista2018evaluating} and Sieve~\cite{mendes2012sieve} enable users to choose metrics among those defined and to declare new ones.
Monitoring tools, such as SPARQLES~\cite{vandenbussche2017sparqles}, also enable assessing some quality aspects.
Instead of these, we choose the IndeGx framework~\cite{maillot2023indegx} because it relies entirely on SPARQL queries, unlike Sieve and Luzzu, and is easily extendable. The IndeGx framework enables querying many KGs, with multiple queries, and storing the results in RDF. Its primary use case is to build an index of KGs and thus to extract and compute various information about them using a SPARQL-based test suite. To evaluate KG accountability, we use it as an engine to submit our own queries to KGs and to store the evaluation results in RDF.

\section{Accountability Requirements and Metric}
\label{sec:definition}
In this section, we define the KGAcc framework. We define the requirements of knowledge graphs accountability, i.e. the precise information that KGs must contain.
To be as unambiguous as possible, we go one step further in expressing these requirements  using SPARQL queries.
Finally, we formally define the metric of accountability.
Our proposal is based on the LiQuID metadata model~\cite{oppold2020accountable} that enables the representation of information related to the accountability of datasets.
To illustrate the use of their model, they provide precise questions that a dataset must answer to be considered accountable.
LiQuID is not specific to any type of dataset, so it is necessary to adapt it.

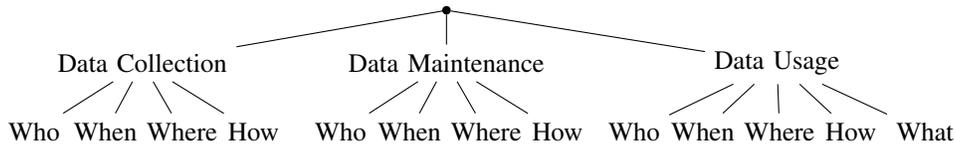
\begin{figure*}[ht!]
\centering
\begin{tikzpicture}[
	scale=1, transform shape,
	level 1/.style={level distance=20pt, sibling distance=115pt},
	level 2/.style={level distance=26pt, sibling distance=28pt},
	child/.style = {align=center}]
  \node [circle,draw,inner sep=1pt,fill=black] {}
    child { node [child] {Data Collection}
      child {node (desc) [child, xshift=1pt] {Who} }
      child {node [child] {When} }
      child {node [child, xshift=1pt] {Where} }
      child {node [child] {How}  }
    }
    child { node [child] {Data Maintenance}
      child {node [child, xshift=2pt] {Who} }
      child {node [child] {When} }
      child {node [child, xshift=1pt] {Where} }
      child {node [child] {How} }
	}
	child[sibling distance=125pt] { node [child] {Data Usage}
      child {node [child, xshift=2pt] {Who} }
      child {node [child] {When} }
      child {node [child, xshift=1pt] {Where} }
      child {node [child] {How} }
      child {node [child] {What} }
    };
\end{tikzpicture}
\caption{The KGAcc Hierarchy of Requirements of Accountability, adapted from LiQuID~\cite{oppold2020accountable} to Fit the Context of Knowledge Graphs}
\label{fig:adim_liquid}
\end{figure*}
\begin{table*}[ht!]
\centering
\caption{Accountability Requirements concerning Data Usage:\\Original Questions from LiQuID and the adapted ones in the KGAcc framework}
\label{tab:question_usage_2}
\begin{small}
\begin{tabular}{|c|p{6cm}|p{6cm}|c|}
    \hline
    & \textbf{Questions from LiQuID} & \textbf{KGAcc Questions} & \textbf{Weight} \\ \hline
    \multirow{3}{*}{\textbf{Usage. Who}} & Who publishes this data set? & Who publishes this KG? & 1 \\ \cline{2-4}
    & \multirow{2}{6cm}{Who has used/ can use the published data set?} & Who has the right to use the published KG? & 1/2 \\ \cline{3-4} 
    & & Who is intended to use the published KG? & 1/2 \\ \hline
    \multirow{3}{*}{\textbf{Usage. When}} & When can/ was the published data set be used? & Since when was the KG available? & 1 \\ \cline{2-4}
    & When is it available? & Until when is the KG available? & 1 \\ \cline{2-4} 
    & Until what point in time is it valid? & Until when is the KG valid? & 1 \\ \hline
    \multirow{3}{*}{\textbf{Usage. Where}} & \multirow{2}{6cm}{Where is the data set published/ available?} & What is the webpage presenting the KG and/or allowing to gain access to it? & 1/2 \\ \cline{3-4}
    & & Where to access the KG (either through a dump or a SPARQL endpoint)? & 1/2 \\ \cline{2-4}
    & Where (place, geographically) can the published data set be used? & In what physical location can the KG be used? & 1 \\ \hline
\end{tabular}
\end{small}
\end{table*}

\subsection{The LiQuID Metadata Model of Accountability}

The LiQuID metadata model relies on a hierarchical structure.
First, it covers all steps of a dataset's life cycle: data collection, processing, maintenance and usage. Then, each life cycle step is structured according to different question types: why, who, when, where, how and what.
Finally, each question type is divided into different fields of information level: description, explanation, legal and ethical considerations and limitations.
The authors provide an exhaustive list of questions to describe each aspect of this hierarchy.
For instance, for ``data processing'', question type ``when'', the question associated with the field ``description'' is ``On what date(s) or time frame(s) has the data been processed?''.
The LiQuID approach proceeds in a very systematic way and requires a large amount of very detailed information, representing what data sources should expose to be as accountable as possible.

\subsection{Adaptation of LiQuID for Knowledge Graphs}

Ideally, to assess the accountability of a KG, we should consider all LiQuID questions. However, it is not possible for all questions to be adapted for KGs and translated into SPARQL queries.

Indeed, as shown by Oppold and Herschel~\cite{oppold2020accountable}, the two general metadata models Dublin Core\footnote{\url{https://dublincore.org/specifications/dublin-core/dcmi-terms/}} and PROV~\cite{prov}, cannot cover all the fields proposed by LiQuID.
According to them, both models ``contain few fields, some of them too general to be mapped to specific LiQuID fields''.
We make the same observation with other general metadata models used in KGs, especially if the task is not to provide the information required by the model, but to query it.
As a consequence of this lack of expressiveness, some questions cannot be translated into queries. As an example, some fields of the information level require too specific information, such as ``Why is it lawful to collect this kind of data?'', which, to our knowledge, cannot be expressed in a KG using existing vocabularies. As another example, two questions result in the same query for different steps of the life cycle, this is in particular the case for the collection and processing steps.

Faced with these difficulties, we opt for a soft strategy in which the maximum score of accountability seems attainable to us.
It consists in keeping only questions compatible with the most common vocabularies of the semantic web.
Therefore, we make the following adaptations: \textit{(i)} only the field ``description'' of the information level is considered, \textit{(ii)} the data processing step of the life cycle level is merged into the data collection step, \textit{(iii)} the question types ``why'', ``what'' in ``data collection'' and ``what'' in ``data maintenance'' are not considered, and \textit{(iv)} two questions concerning the exact methods and tools used for creation and maintenance are not considered in favor of more flexible questions concerning the methodology or procedure only. The resulting hierarchy is shown in Figure~\ref{fig:adim_liquid}. As for the rest of the paper, we omit the last level, as it only contains the ``description'' element.

This definition of the accountability requirements, guided by the desire to ask reasonable questions, leads to a core set of 23 LiQuID questions (out of 207).
We then define the KGAcc requirements by adapting these questions to the context of KGs. We make them more precise, and divide them into smaller parts, so they focus on only one element each. This precision is made as faithfully as possible, with the aforementioned limitations. Table~\ref{tab:question_usage_2} illustrates this adaptation.
Therefore, the KGAcc framework results in 30 questions: 5 for Data Collection, 5 for Data Maintenance, and 20 for Data Usage. The totality of the questions is available on GitHub\footnote{\url{https://github.com/Jendersen/KG_accountability/tree/v2.0/docs}}.

\subsection{SPARQL Implementation of the Questions}
\label{sec:implementation}

Once the questions have been defined, each of them is translated into a SPARQL query or a succession of SPARQL queries. We use more than ten vocabularies of reference, chosen regarding their relevance to describe datasets and concepts around: VoID~\cite{alexander2011describing} is used to express metadata about RDF datasets. DCAT\footnote{\url{https://www.w3.org/ns/dcat}} and DataID\footnote{\url{http://dataid.dbpedia.org/ns/core}} allow the description of datasets and catalogs of datasets. SPARQL-SD~\cite{williams2013sparql} enables to describe SPARQL endpoints. These vocabularies rely on other general vocabularies, the Dublin Core, FOAF\footnote{\url{http://xmlns.com/foaf/spec/}} and SKOS\footnote{\url{https://www.w3.org/TR/skos-reference/}}. We also use PROV-O and PAV~\cite{ciccarese2013pav} for provenance issues. DQV\footnote{\url{https://www.w3.org/TR/vocab-dqv/}} is used to describe the quality of datasets. Finally, we use \href{https://www.schema.org/}{schema.org} a very general and widely used vocabulary, and some specific vocabularies for licenses, such as Creative Commons\footnote{\url{http://creativecommons.org/ns}}.
Each query uses all coherent properties and classes of these vocabularies to be as complete as possible. 
Listing~\ref{lst:publisher_query} shows an example of a question translated into a query, where \verb|?kg| must be replaced by the IRI of the knowledge graph at hand.

\begin{lstlisting}[language=SPARQL, basicstyle=\ttfamily\small, keywordstyle=\color{blue}, backgroundcolor=\color{gray!10}, caption=Extended query associated with ``Who publishes this dataset?'', label=lst:publisher_query]
PREFIX dct: <http://purl.org/dc/terms/>
PREFIX dce: <http://purl.org/dc/elements/1.1/>
PREFIX schema: <http://schema.org/>
PREFIX prov: <http://www.w3.org/ns/prov#>
ASK {
 {?kg dct:publisher ?publisher .}
 UNION {?kg dce:publisher ?publisher .}
 UNION {?kg schema:publisher ?publisher .}
 UNION {?kg schema:sdPublisher ?publisher .}
 UNION {?kg prov:wasGeneratedBy ?act .
   ?act a prov:Publish .
   ?act prov:wasAssociatedWith ?publisher .} }
\end{lstlisting}

As queries are associated with questions requiring answers, they are ``ASK'' queries. The answer TRUE is considered a success, as it means that the KG contains the desired information.
On the opposite, the answer FALSE or an error (e.g., timeout exception) is a failure, as it means the KG is unable to provide the wanted information.

Finally, notice that to express our precise requirements, we can either use the queries in their extended version, including all possible ways of expressing the required information, as in Listing~\ref{lst:publisher_query}.
Alternatively, we can express the requirements in the form of a compact query, as in Listing~\ref{lst:simple_publisher}, completed with a set of equivalences between properties (and between more complex graph patterns if necessary).

\begin{lstlisting}[language=SPARQL, basicstyle=\ttfamily\small, keywordstyle=\color{blue}, backgroundcolor=\color{gray!10}, caption=Compact query associated with ``Who publishes this dataset?'', label=lst:simple_publisher]
PREFIX dct: <http://purl.org/dc/terms/>
ASK { ?kg dct:publisher ?publisher . }
\end{lstlisting}

\subsection{Definition of the Metric}
\label{sec:measure}

First, we define the score obtained for each question.
Then it is possible to determine the score of each node of the KGAcc hierarchy defined in Figure~\ref{fig:adim_liquid}, from the bottom to the top. The score at the top of the hierarchy is the overall accountability score.

A successful query gives a score of 1 to its associated question, while a failure gives 0.
The score of a question associated with a succession of queries is the average of the score given by each query.
The accountability score of a leaf of the KGAcc hierarchy (e.g. ``data usage - who'') is the weighted average of the scores obtained for its associated questions.
Notice that LiQuID does not weight its questions, which suggests they are of equal importance. To stay close to this, we use to the following rule. When $m$ ($m\geq 1$) KGAcc questions come from a same LiQuID question, a weight of $1/m$ is associated to each of them.
Table~\ref{tab:question_usage_2} illustrates these weights.
For instance, ``data usage - who'' has three questions, coming from two LiQuID questions. The first leads to one question, so its weight is $1$, and the second leads to two questions, therefore their weight is $1/2$ each. The accountability score of  ``data usage - who'' is the weighted average of these three questions, with the weights of $1$, $1/2$, and $1/2$.
For the other elements of the hierarchy, we determine their score by computing the (non-weighted) average of the scores of the elements underneath.

Formally, let $g$ be a knowledge graph, $\ell$ a leaf node of the KGAcc hierarchy (e.g. ``data usage - who''), and let $Q(\ell)$ denote all questions associated with $\ell$. With $\mathit{score}$ a function giving the score of $g$ for a given question $q$, $w_q$ the weight of question $q$, the accountability score of $g$ w.r.t. $\ell$ is:
\begin{equation}
\mathit{accountability}(g, \ell) = \frac{\sum\limits_{q \in Q(\ell)} w_q \cdot \mathit{score}(g, q)}{\sum\limits_{q \in Q(\ell)} w_q} 
\end{equation}
and the score of a given node $n$ of the KGAcc hierarchy which is not a leaf is:
\begin{equation}
\mathit{accountability}(g, n) =  \frac{\sum\limits_{n' \mathit{ child}_{} \mathit{of}_{} n} \mathit{accountability}(g, n')}{\mathit{\ number\ of\ children\ of\ } n}
\end{equation}
In particular, the global accountability score is the score for the upper node in the hierarchy.

\section{Experimentation and Results}
\label{sec:results}
In this section, we describe our experiments. First, we detail the method employed to conduct an evaluation campaign of several KGs. Then, we discuss two aspects of the results. First, we examine the capabilities of knowledge graphs with regard to accountability.
Second, we discuss the measure itself and the relevance of the KGAcc questions.
All our queries and results are publicly available on our GitHub repository\footnote{\url{https://github.com/Jendersen/KG_accountability/tree/v2.0}}.

\subsection{Processing and Tool}
To evaluate a knowledge graph, we first need to identify its IRI within its own data.
Then, we proceed in several stages to evaluate how the KG answers the queries defined in the previous section.

An important prerequisite of all our queries is to identify the IRI that the studied KG uses to refer to itself, or more precisely, the IRIs of the datasets it contains.
Indeed, this IRI  is the subject of at least one triple in all our queries, as illustrated in Listing~\ref{lst:publisher_query}.
Therefore, a query looking for the IRI is defined and presented in Listing \ref{lst:iri_kg}, where \verb|$rawEndpointUrl| is replaced by the URL of the endpoint during evaluation. If the KG does not provide an answer to this query, it will not answer any of our queries.

\begin{lstlisting}[language=SPARQL, basicstyle=\ttfamily\small, keywordstyle=\color{blue}, backgroundcolor=\color{gray!10}, caption=Query to identify the IRI of the studied KG, label=lst:iri_kg]
SELECT ?kg WHERE {
  ?kg ?endpointLink $rawEndpointUrl .
  { ?kg a dcat:Dataset }
  UNION { ?kg a void:Dataset }
  UNION { ?kg a dcmitype:Dataset }
  UNION { ?kg a schema:Dataset }
  UNION { ?kg a sd:Dataset }
  UNION { ?kg a dataid:Dataset }
}
\end{lstlisting}

In order to reduce the complexity of the queries sent to KGs, we focus on the accountability requirements in their compact form (cf. Listing~\ref{lst:simple_publisher}).
So, we proceed according to the following steps.
For each KG, we first extract the triples corresponding to its metadata.
As explained before, we use queries that begin with the lines described in Listing~\ref{lst:iri_kg} to identify them.
Then, we saturate this description of the KG by adding equivalent properties and classes, as defined by the requirements.
Finally, we evaluate a KG according to this saturated metadata, using compact queries.

To carry out all these steps, we use the framework IndeGx\footnote{\url{https://github.com/Wimmics/dekalog}}.
It relies on a SPARQL-based test suite and can pre-process some steps for a better scalability.
To use it, we provide SPARQL queries and configure the actions to be taken based on their results, i.e. which triples to write or update in the resulting RDF graph.
So, we embed a set of queries into the framework, following the steps previously detailed, and declare how to store the result (True or False) for each evaluation query and for each KG using the DQV Vocabulary.

\subsection{Querying of the SPARQL Endpoints}

Our experiments query the 336 SPARQL endpoints already identified by IndeGx, extracted from LOD Cloud, Wikidata, SPARQLES, Yummy Data, and Linked Wiki in February 2023.
These endpoints were queried at three different time points in June 2023.
For each endpoint, only the results of an experiment for which it was available are kept. In this way, KGs are not penalized if they were unavailable at a given time.
All endpoints not succeeding the query of Listing~\ref{lst:iri_kg} are assigned an accountability score of 0, as no triple concerning its KG could be extracted.

Finally, given the results obtained for each query and thus each question, the accountability score can be computed. As defined in subsection~\ref{sec:measure}, an average is used to calculate the score for each aspect of the KGAcc hierarchy of Figure~\ref{fig:adim_liquid}, until the overall accountability score is obtained.

\subsection{Analysis of the Results}

Among the 336 endpoints tested, only 26 successfully provide accountability information (Listing \ref{lst:iri_kg}). The others were unavailable or did not provide easily accessible meta-information within their data.
Among the 26 endpoints, 166 different datasets were identified (in the sense of \verb|dcat:Dataset| or \verb|void:Dataset|\ldots), with accountability scores varying between 3.1\% and 59\%, with an average score of 26\%.
Even though most of the KGs do not provide any accountability information, the distribution of the values shows that this measure allows to discriminate between the datasets and the 26 endpoints left.
On average, KGs are more accountable concerning ``data usage'' (41\%) than ``data collection'' (25\%), and twice better on ``data collection'' than on ``data maintenance'' (12\%).

\begin{figure}[ht!]
\centering
\includegraphics[scale=0.6]{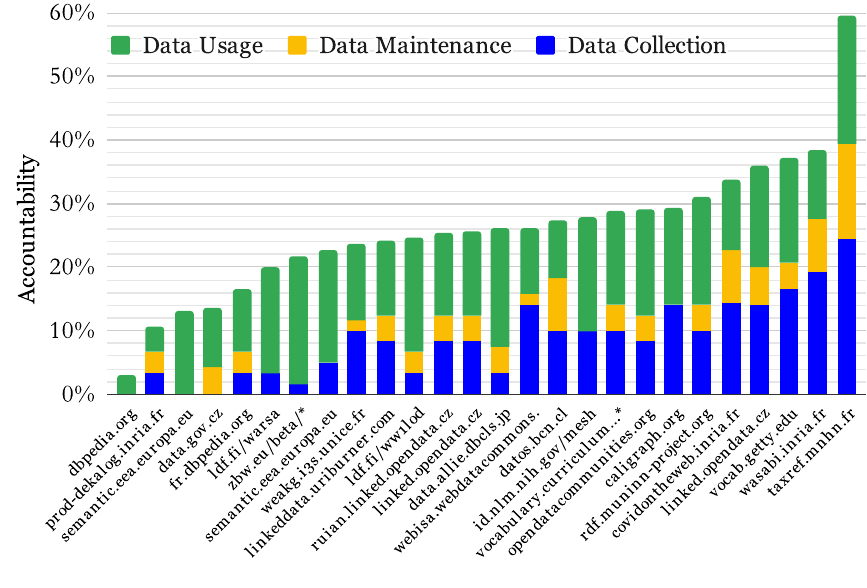}
\begin{small}
All URLs start with \url{http(s)://} and, except for *, end with \url{/sparql}.
\end{small}
\caption{Accountability Score Obtained by the Best Dataset of Each Evaluated Endpoint, Detailed according to the Three Life Cycle Steps.}
\label{fig:Results_KG}
\end{figure}

\begin{figure}[ht!]
\centering
\scalebox{0.6}{\input{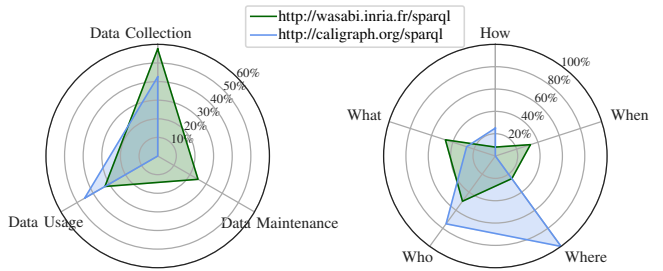}}
\centering
\subfloat[][Accountability of Two KGs w.r.t. the Life Cycle Steps]{%
	\hspace{0.23\textwidth} \label{fig:Radar:lifecycle}
}%
\hfill
\subfloat[][Accountability of Two KGs w.r.t. the Question Types of ``Data Usage'']{\hspace{0.23\textwidth} \label{fig:Radar:usage}}

\caption{Accountability of Two KGs w.r.t. Different Elements of the Hierarchy}
\label{fig:Radar}
\end{figure}

Figure \ref{fig:Results_KG} shows the accountability score of the best dataset of each endpoint. This score is divided according to the three life cycle steps ``data collection’’, ``data maintenance’’ and ``data usage’’.
It is possible to compare two datasets in more detail. As an example, Figure \ref{fig:Radar} shows the strengths and weaknesses of the main dataset of \url{http://caligraph.org/sparql} and of \url{http://wasabi.inria.fr/sparql} according to the life cycle steps (\ref{fig:Radar:lifecycle}) and more precisely on the question types of the ``data usage'' step (\ref{fig:Radar:usage}).

The small number of KGs having a non-zero accountability score is not surprising. This observation is in line with other results~\cite{maillot:hal-03652865} showing that less than 10\% of KGs provide self-descriptions within their data. 
However, for some KGs, it is possible that some meta-information may be present outside of the KG itself, for instance on their web page, or inside the KG but not findable in the way shown in Listing~\ref{lst:iri_kg}. While not taking them into account may penalize some KGs, it points out the fact that they are less transparent because the information is less accessible.

Several reasons may explain the scores obtained on the different life cycle steps. 
The ``data usage'' step covers general description elements that are widely used such as a description, a publisher, or a license, that more than half of the 26 endpoints provide. Furthermore, it encompasses all questions involving VoID vocabulary, which are each answered at least once by more than 50\% of the endpoints on average.
``Data usage'' also requires a link to the endpoint or a dump, so having an answer to this question is expected considering how we identify the IRI of the KG.
``Data maintenance'' usually has bad scores. This may be due to the fact that half of the questions have only one possible property, with no alternative. For instance, the modification frequency can only be obtained with the property \verb|accrualPeriodicity| from the Dublin Core vocabulary.
This lack of alternative solutions to express this concept makes it more difficult to answer the query and highlights the fact that the question is more unusual.
``Data creation'' has various results with very common requirements, such as the creator and the creation date, and more difficult questions to answer such as the creation method.

\subsection{Discussion about the KGAcc Framework}
\begin{figure}[ht!]
\centering
\scalebox{0.71}{\input{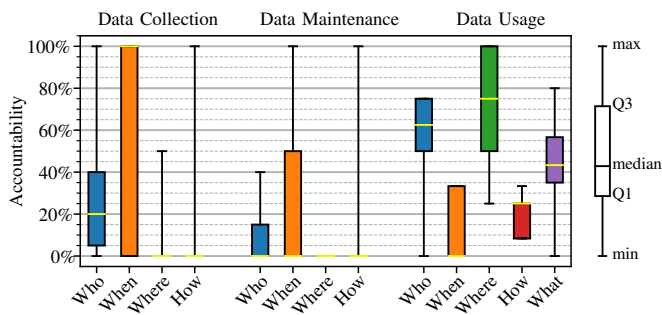}}
\caption{Box Plot Showing the Distribution of the Values of Accountability w.r.t. Each Aspect of the Hierarchy.}
\label{fig:Results_KG_moustache}
\end{figure}

As far as the framework is concerned, at least two questions can be asked: are the requirements relevant? Are they too demanding? Figure~\ref{fig:Results_KG_moustache} provides some answers. It represents the distribution of the values of accountability of the best dataset of each endpoint w.r.t. each aspect of the KGAcc hierarchy. Each box represents the first quartile (Q1), the median, and the third quartile (Q3) of the values obtained on the different aspects and the whiskers indicate the minimum and the maximum values obtained.
It shows that the question types of ``data usage'' usually have good values in the different KGs. It also shows that half of the aspects can be fully covered, including ``data collection - who'', ``data collection - when'' and ``data collection - how'' for instance.

On the one hand, as many of the aspects sometimes get the maximum score, it shows that these queries are relevant and that KGs have a good margin to improve themselves.
On the other hand, some of the low scores observed on that figure may be explained by too demanding queries. Indeed, it is important to notice that 6 out of 30 queries never succeed. For instance, in ``data usage - when'' the end date of availability of the dataset may be difficult for providers to specify, as they may consider that their KGs will be available indefinitely.
Other questions concerning locations may not be in line with the current practices. Indeed, they are especially important for KGs that hold private information, which is generally not the case for public SPARQL endpoints.
As other frameworks, depending on the context, KGAcc may be discussed and improved with experts and KG providers.

\section{Comparison with some Evaluation Frameworks}
\label{sec:comparison}

To take the analysis of our framework a step further, we compare it in detail with several data quality and FAIRness assessment frameworks. The comparison is made at the level of the required properties: \textbf{we aim to verify to what extent other studies require the properties demanded by the KGAcc framework}. To do so, we focus on studies that consider RDF properties and RDF datasets, and that either provide an open access implementation or that describe the metrics in sufficient detail to allow comparison. This is why works such as F-UJI~\cite{devaraju2021automated} or Sieve~\cite{mendes2012sieve} are not considered.

Concerning data quality, Zaveri et al.~\cite{zaveri2016quality} provide an organized list of metrics obtained by a systematic literature review. This work is theoretical and does not implement these metrics, therefore, we do not take it into account.
However, we focus on two major studies of data quality inspired by Zaveri et al.
First, Färber et al.~\cite{farber2018linked} evaluate the data quality of five cross-domain KGs, namely DBpedia, Freebase, OpenCyc, Wikidata, and YAGO. While the implementation of their metrics is not available, each metric is richly described.
Secondly, Debattista et al.~\cite{debattista2018evaluating} provide a more generic set of data quality metrics enabling the evaluation of any KG. Their implementation is available online but the article~\cite{debattista2018evaluating} describing them is more detailed and understandable.
Therefore, for all these studies, we base our comparison solely on the referenced article.

For FAIRness, FAIR-checker~\cite{rosnetfair} is interested in RDF triples as embedded metadata in web pages. For comparison, we rely on the specifications provided by the online tool\footnote{\url{fair-checker.france-bioinformatique.fr/check} Accessed: 10 April 2023} when evaluating a resource.
We also consider O'FAIRe~\cite{amdouni2022faire} which focuses on RDF ontologies. It provides an online tool\footnote{\url{agroportal.lirmm.fr/landscape\#fairness_assessment}} to see the results obtained by a list of ontologies. To compare with them, we consider the complete list of questions and their required properties\footnote{\url{https://github.com/agroportal/fairness/blob/master/doc/results/FAIR-questions.md} Accessed: 10 April 2023}.
In total, this leads us to consider two data quality studies and two FAIRness ones.

\begin{table*}
\centering
\caption{Comparison Between the KGAcc Queries and the Metrics Proposed by the Works on Data Quality and FAIR}
\label{tab:comparison}
\include{liquid/comparison_table.tex}
\end{table*}

Table~\ref{tab:comparison} summarizes our comparative study.
For each KGAcc query, if one of its required properties is also required in a data quality or FAIR metric, a mark is indicated in the table. If this property must not necessarily concern the KG (e.g. the creator of a resource instead of the creator of the dataset), then the mark is $\approx$, showing that the FAIR or data quality metric is not really related to dataset accountability. Otherwise, if this property is mandatory to obtain a maximum score on the data quality or FAIR metric, the mark is $\checkmark$. If the property is listed among other properties and only one or two or $n$ of these properties are necessary for success, then the mark is $\subset$. For instance, in O'FAIRe the third question for principle F2 states that to obtain the maximum score, six properties should be used from a list of 37 properties, whatever those six properties may be. Therefore, unlike the $\checkmark$ mark, a $\subset$ mark does not guarantee that passing the FAIR metric ensures passing the accountability query.

This table highlights several elements concerning data quality metrics.
First, as part of the measure of accessibility, all these evaluations require a license to be present using properties such as \verb|dcterms:licence|~\cite{farber2018linked,debattista2018evaluating}.
They also demand some particular provenance information: the creators or the publishers of the KG~\cite{debattista2018evaluating}, or other information not specifically related to the KG such as the source of some data to enhance trustworthiness~\cite{farber2018linked}, their modification dates~\cite{farber2018linked}, or traceability of the data~\cite{debattista2018evaluating}.
Some other meta-information is expected to be provided, such as the serialization formats~\cite{debattista2018evaluating}.
Finally, Färber et al.~\cite{farber2018linked} request the provision of KG metadata citing as an example the URI of the SPARQL endpoint or the RDF export URL to indicate where to access the data.

Concerning FAIR metrics, only two metrics are related to accountability in FAIR-checker. The first one measures the `R1.1' principle that requires a license. The second one measures the provisioning of provenance information (R1.2) by checking that at least one of the 23 listed properties is found (such as \verb|prov:wasDerivedFrom|, \verb|pav:createdBy|, etc.). O'FAIRe offers more similarities with our work. There are mainly two different kinds of metrics of interest compared to our queries. First, the metrics concerning the reusability principles focus on one information each and are mostly also required by accountability (creator, contributor, source, method, periodicity, license and rights). Secondly, some metrics of findability require that the ontology uses some well-known properties. Indeed, two questions of the principle `F2' cover properties required by at least 11 accountability queries (such as \verb|dct:created|, \verb|void:dataDump|...).

As a result, both data quality and FAIRness share common interests with accountability.
Therefore, improving them may have a positive impact on the assessment of accountability and vice versa.
However, neither data quality nor FAIRness focuses specifically on accountability as a whole and does not take into account all the elements it requires.
The general studies on data quality only slightly overlap with accountability.
FAIRness has more similarities, particularly with regard to the steps of data collection and maintenance as it is mainly interested in questions of provenance.
In particular, O'FAIRe seems to have many similarities.
However, most of them result solely from the two findability metrics, which are not very informative about the type of metadata present, since they cover no more than 11 of our queries.
Therefore, the measure of accountability is much more detailed, precise, and focused than O'FAIRe on our point of interest. And as the result of each query is available, the former provides a much more relevant view of accountability.

\section{Conclusion}
\label{sec:ccl}
In order to evaluate the accountability of RDF graphs, we proposed : \textit{(i)} the KGAcc framework defining requirements concerning the metadata that the KGs should expose and an associated metric, \textit{(ii)} the evaluation of a large set of endpoints, \textit{(iii)} a comparison of our approach with other frameworks that assess data quality or compliance to the FAIR principles.

The KGAcc requirements are expressed as SPARQL queries. They are obtained through a meticulous adaptation of an existing hierarchy of natural language questions, proposed for datasets in general~\cite{oppold2020accountable}, to the specific context of KGs. Indeed, the dedicated vocabularies make easy stating some requirements. But their lack of expressiveness makes some questions collapse into a same query or prevents from considering demanding and precise questions about ethical and legal questions. This is why we end up with a relatively small set of queries.

The evaluation of many RDF graphs reveals that most of them do not provide any of the required information within their data, even though some of the required information is very commonly requested.
However, there are RDF graphs that provide some of the expected information, showing our demands are reasonable.
Our comparative study shows in particular that O'FAIRe is the framework considering the most properties common to accountability. However, it is not so demanding in terms of accountability and cannot be considered as a framework dedicated to this aspect.

In future works, to improve our measure, we will introduce some weights to aggregate the results differently, and we will propose an online visualization of the results.
Finally, it could be interesting to automate some tasks, such as defining the equivalences.

\bibliographystyle{IEEEtran}
\bibliography{main.bib}

\end{document}

%% file: liquid/comparison_table.tex
\begin{tabular}{|c|c|l|c|c|c|c|}
\hline
\textbf{Lifecycle} & \textbf{Question} & \multicolumn{1}{c|}{\textbf{Accountability}} & \multicolumn{2}{c|}{\textbf{Data Quality}} & \multicolumn{2}{c|}{\textbf{FAIR}} \\ \cline{4-7} 
\textbf{step} & \textbf{type} & \multicolumn{1}{c|}{\textbf{query}} & \textbf{Debattista}~\cite{debattista2018evaluating} & \textbf{Färber}~\cite{farber2018linked} & \textbf{FAIR-Checker}~\cite{rosnetfair} & \textbf{O'FAIRe}~\cite{amdouni2022faire} \\ \hline \hline

\multirow{6}{*}{\begin{tabular}[c]{@{}c@{}}Data\\ Collection\end{tabular}} & \multirow{2}{*}{Who} & Creator & $\checkmark$ &  & $\subset$ & $\checkmark$ \\ \cline{3-7} 
 &  & - Creator's info. &  &  &  &  \\ \cline{2-7} 
 & When & Creation date &  &  & $\subset$ & $\subset$ \\ \cline{2-7} 
 & \multirow{2}{*}{Where} & Source &  & $\approx$ & $\subset$ & $\checkmark$ \\ \cline{3-7} 
 &  & Creation location &  &  & $\subset$ &  \\ \cline{2-7} 
 & How & Creation method &  &  & $\subset$ & $\checkmark$ \\ \hline \hline
 
\multirow{6}{*}{\begin{tabular}[c]{@{}c@{}}Data\\ Maintenance\end{tabular}} & \multirow{2}{*}{Who} & Contributor &  &  & $\subset$ & $\checkmark$  \\ \cline{3-7} 
 &  & - Contributor's info. &  &  &  &  \\ \cline{2-7} 
 & \multirow{2}{*}{When} & Modification date &  & $\approx$ & $\subset$ & $\subset$  \\ \cline{3-7} 
 &  & Frequency &  &  &  & $\checkmark$  \\ \cline{2-7} 
 & Where & Modification location &  &  &  &  \\ \cline{2-7} 
 & How & Modification method &  &  & $\subset$ &  \\ \hline \hline
 
\multirow{21}{*}{\begin{tabular}[c]{@{}c@{}}Data\\ Usage\end{tabular}} & \multirow{3}{*}{Who} & Publishers & $\checkmark$ &  &  & $\subset$ \\ \cline{3-7} 
 &  & Usage rights & $\checkmark$ & $\checkmark$ & $\checkmark$ & $\checkmark$ \\ \cline{3-7} 
 &  & Audience &  &  &  & $\subset$  \\ \cline{2-7} 
 & \multirow{3}{*}{When} & Start of availability &  &  &  &   \\ \cline{3-7} 
 &  & End of availability &  &  &  &   \\ \cline{3-7} 
 &  & End of validity &  &  &  & $\subset$ \\ \cline{2-7} 
 & \multirow{3}{*}{Where} & Webpage &  &  &  & $\subset$ \\ \cline{3-7} 
 &  & Access URL &  & $\checkmark$ &  & $\subset$ \\ \cline{3-7} 
 &  & Usage location & $\checkmark$ & $\checkmark$ & $\checkmark$ & $\checkmark$  \\ \cline{2-7} 
 & \multirow{5}{*}{How} & License & $\checkmark$ & $\checkmark$ & $\checkmark$ & $\checkmark$ \\ \cline{3-7} 
 &  & Access URL &  & $\checkmark$ &  & $\subset$ \\ \cline{3-7} 
 &  & - Endpoint's info. &  &  &  &  \\ \cline{3-7} 
 &  & Usage information &  &  &  &  \\ \cline{3-7} 
 &  & Usage requirements &  &  &  &  \\ \cline{2-7} 
 & \multirow{7}{*}{What} & Examples &  &  &  &  \\ \cline{3-7} 
 &  & Concepts &  &  &  & $\subset$ \\ \cline{3-7} 
 &  & Description &  &  &  & $\subset$ \\ \cline{3-7} 
 &  & Triples &  &  &  &   \\ \cline{3-7} 
 &  & Entities prop. classes &  &  &  &   \\ \cline{3-7} 
 &  & Serialization & $\checkmark$ &  &  &   \\ \cline{3-7} 
 &  & Quality &  &  &  & $\subset$ \\ \hline
 
\multicolumn{7}{l}{\quad $\checkmark$ \quad Property required and must concern the dataset (or the ontology).} \\
\multicolumn{7}{l}{\quad $\approx$ \quad Property required but not linked to any particular resource.} \\
\multicolumn{7}{l}{\quad $\subset$ \quad One of the required properties is listed in the metric among other semantically different properties.}
\end{tabular}